# Unlocking the Potential of Open Government Data: Exploring the Strategic, Technical, and Application Perspectives of High-Value Datasets Opening in Taiwan

Hsien-Lee Tseng

Department of Public Administration and Management, National University of Tainan, Taiwan, nova1219@gmail.com

Anastasija Nikiforova

Institute of Computer Science, Faculty of Science and Technology, University of Tartu, Estonia,
nikiforova.anastasija@gmail.com

**Abstract**: Today, data has an unprecedented value as it forms the basis for data-driven decision-making, including serving as an input for AI models, where the latter is highly dependent on the availability of the data. However, availability of data in an open data format creates a little added value, where the value of these data, i.e., their relevance to the real needs of the end user, is key. This is where the concept of "high-value dataset" (HVD) comes into play, which has become popular in recent years. Defining and opening HVD is an ongoing process consisting of a set of interrelated steps, the implementation of which may vary from one country or region to another. Therefore, there has recently been a call to conduct research in a country or region setting considered to be of greatest national value. So far, only a few studies have been conducted at the regional or national level, most of which consider only one step of the process, such as identifying HVD or measuring their impact. With this study, we answer this call and examine the national case of Taiwan by exploring the entire lifecycle of HVD opening. The aim of the paper is to understand and evaluate the lifecycle of high-value dataset publishing in one of the world's leading producers of information and communication technology (ICT) products - Taiwan. To do this, we conduct a qualitative study with exploratory interviews with representatives from government agencies in Taiwan responsible for HVD opening, exploring HVD opening lifecycle. As such, we examine (1) strategic aspects related to the HVD determination process, (2) technical aspects, and (3) application aspects.

## 1 Introduction

Today, data has an unprecedented value as data is the basis for an informed data-driven decision-making, including serving as an input for Artificial Intelligence (AI) models, where the latter is highly dependent on the availability, quality, and suitability of the data for the intended purpose. In other words, "*Data is the*

*Lifeblood of AI*" [1]. This is even more relevant in todays' trend towards open AI - democratised and more accessible than ever, where a competitive advantage has those who have data to fuel AI. Data democratisation is another trend running alongside the democratisation of AI, where open data and open government data (OGD) have been an ongoing trend for several decades. However, availability of data in an open data format creates a little added value, where the value of these data, i.e., their relevance to the real needs of the end user, is key. This is where the concept of "high-value dataset" (HVD) comes into play, which has become popular in recent years.

Although this topic has been studied over years, it has gained greater popularity due to the progress made by the European Commission in the Open Data Directive, which defined and subsequently detailed thematic data categories of HVD, including technical requirements that they must meet for an increased harmonization and interoperability of public sector data in EU Member States, as well as their overall readiness for reuse and compliance with expected macro-characteristics, which include *(1) economic benefits, (2) environmental benefits, (3) social benefits, (4) generation of innovative services and innovation* becoming of more importance with the development of advanced digital technologies, such as artificial intelligence, distributed ledger technologies (DLT) and the internet of things (IoT), *(5) reuse, and (6) the improvement, strengthening and support of public authorities* in carrying out their missions (public services and public administration, social) [2].

However, defining and opening HVD is a process consisting of a set of interrelated steps, the implementation of which may vary from one country or region to another. Therefore, there has recently been a call to conduct research in a country or region setting considered to be of greatest national value [3]. So far, only a few studies have been conducted at the regional or national level, namely Croatia [4], Latvia [5], the Netherlands [6], and Thailand [7]. Moreover, most of them consider only one step of the process, such as identifying HVD or measuring their impact. With this study, we answer this call and examine the national case of Taiwan by exploring the entire lifecycle of HVD opening.

Although Taiwan OGD is considered competitive and one of the leading OGD initiatives in the world, e.g., Taiwan ranks 6th out of 63 economies in the World Competitiveness Yearbook in 2023, and 11th in Digital Competitiveness ranking [8], it is rarely covered in the literature. Most studies analyse (1) user intentions to use OGD by users [9], or (2) government agencies' data sharing intentions using one of the technology adoption theories [10, 11].

The OGD movement began in Taiwan in 2012 with the launch of the OGD portal, designed to integrate open data from various agencies and local governments. Although the portal has evolved over the years, according to *Taiwan Open Government National Action Plan 2021-2024*, the private sector would prefer a more comprehensive system and mechanism to optimize the value of using government open data. Therefore, the first of five commitments of *Taiwan Open Government National Action Plan 2021-2024* [12] concerns the provision of high-value datasets. It consists of three commitments: (1) to focus on prioritising high value data opening, (2) strengthen data standards and format quality, and (3) establish processes for meeting public data needs. As part of the Action Plan six (6) topics that represent the areas of HVD were identified, namely: *(1) Climate and Environment, (2) Disaster Prevention and Relief, (3) Transportation and Transit, (4) Healthcare and Medical Services, (5) Energy Management, and (6) Social Assistance*. Within these categories, specific datasets were identified as HVD and assigned to respective Taiwanese agencies to be opened in 2022.

The aim of the paper is to understand and evaluate the publication lifecycle of high-value datasets in Taiwan. To do this, we consider three perspectives: (1) **a strategic perspective** associated with the HVD determination process, (2) **a technical perspective** related to the dataset preparation stage, and (3) **an application perspective** associated with the further assessment of the generated impact by the published HVD re-use.

In more detail, the strategic perspective includes establishing an understanding of *(1a) how the HVD are determined, (1b) what the time cycle of the aforementioned process is, (1c) what the rewards or other incentives are for stakeholders involved in the discovery, opening and promotion of HVDs or is it rather a compulsory effort, determined by law or policy, without further support from the public agency involved in their opening process?* The technical perspective examines *(2a) the methods and channels used or required to assess needs when selecting HVD, as well as establishes (2b) whether the requirements for HVD differ from those imposed for general open data, referring to aspects such as data quality, granularity, update frequency, integration methods etc.* Finally, we find out *(2c) is there a plan to include non-personal data sets in HVD.* The application perspective focuses on the impact created by published HVD. First, we determine *(3a) whether regular follow-up monitoring and analysis of applications and services developed from HVDs, especially those that have social or environmental value, e.g., in areas such as weather, traffic, net-zero initiatives*. We then determine (3c) *how the impact of HVD is assessed (and whether this is done?)*

To attain the objective of this study, a semi-structured interview instrument was developed to collect empirical data to understand the current state of HVD opening in Taiwan. Four (4) ministries - *The Ministry of Digital Affairs, Ministry of the Interior, Department of Statistics, Ministry of Transportation and Communications, Ministry of Environment, Department of Monitoring and Information*, constitute the sample for this study responsible for HVD opening and therefore have extensive experience in implementing respective open data policies.

Considering Taiwan as a use-case can provide a better understanding of the current state of affairs in Taiwan, thereby creating national value for the country in question, filling the gap of decentralised and dispersed/ scattered understanding even within the country, and also identifying weaknesses, which need to be improved to further more successfully open and maintain HVD, thereby contributing to the maturity and sustainability of Taiwan's open data ecosystem, as well as to provide insights that could be of potential value to other countries.

The rest of the paper is structured as follows: Section 2 provides a research background, Section 3 presents the methodology of the research, Section 4 presents the case study, i.e., its summary and findings, while Section 5 and 6 establish the discussion and elaborates on limitations, while Section 7 concludes the paper.

## 2 Background

In this section, we provide the background on (1) the current state of HVD in individual country contexts, referring to both practical developments and the scientific literature, and (2) the state of affairs in the HVD area in Taiwan.

### 2.1 HVD in the context of individual country or region

While the EU's efforts can be seen as the most advanced and promising in Europe, it should be acknowledged that these efforts are not the only ones, and there are other independent initiatives to identify the most valuable datasets for a given region or country. They are often older than those promoted with the OD Directive.

For example, according to [13], in 2016, data.overheid.nl (the **Netherlands**) in collaboration with municipalities, the Digital Urban Agenda and VNG/KING, compiled a Municipal High Value List to provide municipalities with support in opening data and assisting to prioritise specific dataset opening. These HVDs belong to the so-called "*Data with impact*" category, which, in addition to HVD consists of "*reference datasets*" and "*nationwide datasets*", with the HVD consisting not only of municipal but also of local HVDs. In the context of HVD, the value of a dataset and its belonging to HVD is determined by the extent to which the data contribute to: *(1) transparency, (2) legal duty, (3) cost reduction, (4) target audience, and (5) potential for reuse*, where the G8 Open Data Charter and the 14 categories it has identified as 'high value' datasets are used as a reference list to indicate potentially HVD that will benefit data owners in opening the most relevant and useful datasets. When compiling national lists of HVD, data.overheid.nl also draws inspiration from other benchmarks, such as the *Open Data Barometer* (from Open Data Institute), the *Open Data Maturity Report*, the *Global Open Data Index* (from the Open Knowledge Foundation), *Country Factsheets as part of the Government at a Glance OECD*, and the *Nederlandse benchmarks* (from 2014 to 2016).

In **Canada**, the work in the identification of high-value datasets has been ongoing for years, with each government jurisdiction establishing its own criteria for identifying HVDs. However, the *Canada Open Government Working Group (COGWG)* was tasked later - as part of the *Third Biennial Plan to the Open Government Partnership (2016-18),* with developing common criteria that would identify HVDs that could be released across jurisdictions to spur innovation and generate greater socio-economic impact [14]. These datasets and corresponding examples of common dataset types were based on jurisdictional scanning of high value dataset criteria, surveys, and the International Open Data Charter. The HVD criteria are, in turn, five, namely: to be able to *(1) help identify social, environmental and economic conditions, (2) helps achieve better outcomes in public services, (3) encourage innovation and sustainable economic growth, (4) increases transparency, accountability and information flow of government,* and *(5) is in high demand from the community*, i.e., meeting the needs and values of citizens, measured through requests for specific datasets by citizens and communities as an indication of public demand for these data.

However, the pioneer in the HVD is seen to be **Australia** [15] and its *First National Action Plan of the Australian Open Government Partnership 2016-2018* [16] that contained among its objectives the implementation of actions to develop and publish a framework for HVDs to *generate new business, develop new products and services*, and *create social value* and to design how best to facilitate the sharing and use of these datasets through the legislative consultation process, which included consultations with representatives of the research, not-for-profit and private sectors to identify HVDs for release. Engaging non-governmental stakeholders - as part of the *Better and More Accessible Digital Services* policy - is necessary to identify the characteristics of HVDs, such as format, *accessibility, metadata* and *attributes*, and *identify barriers to data access and sharing*. As part of this, the Government established *Smart Cities and Suburbs Program* to encourage local councils to open their data and collaborate with communities, local businesses, not-for-profit and research institutes to create innovative solutions to urban problems. However, in 2019, a delay to the initiative was announced, and although work was announced to resume later, no documentation on further progress was found (also in line with [15]).

Five years ago, in 2019, the term "high-value dataset" was also introduced by the European Parliament and the Council of the **European Union**, marking the start of efforts to determine and promote HVD in the EU Member States. The first document - *Directive (EU) 2019/1024 of the European Parliament and of the Council of 20 June 2019 on open data and the re-use of public sector information* [17], defined six thematic data categories of HVD – *(1) geospatial, (2) earth observation and environment, (3) meteorological, (4) statistics, (5) companies and company ownership, (6) mobility*. These categories were subsequently detailed in December 2022 in *Commission Implementing Regulation (EU) 2023/138* [18], which Member States must apply by June 2024. Committed to greater harmonization and interoperability of public sector data across EU Member States, it defined *(1) specific datasets, their (2) granularity, (3) key attributes, (4) geographical coverage, (5) requirements for their re-use, including (5.1) licence, (5.2) specific format where appropriate, (5.3) frequency of updates and timeliness, (5.4) availability in a machine-readable format, (5.5) accessibility via API and bulk download, (5.6) support for metadata* describing data within INSPIRE data themes, which must contain a certain minimum set of the required metadata elements, description structure and semantics of the data, use of controlled vocabularies and taxonomies etc.

According to the recent Open Data Maturity Report (ODM) [19], Member States are actively preparing to open high-value datasets, with 25 Member States working to implement Commission Implementing Regulation (EU) 2023/138, with **Estonia**, **Finland**, **Denmark**, **Latvia**, **Czechia** and **Slovenia** found significantly ahead of other EU Member States in preparing for HVD (according to data these countries reported on themselves). 22 Member States are also actively working to ensure the interoperability of these datasets. According to the ODM report 2023, category-wise, *geospatial* and *statistics* datasets are seeing the most progress, while other HVD are opened up less actively. While from a HVD lifecycle perspective, *identification* and *inventory*, as well as addressing *legal barriers* are tasks performed by more countries and at a higher level, with the lowest scores for *meeting technical requirements*, such as *metadata quality*, *standardised structures*, *access* and *machine-readable formats* showing the lowest average levels of progress.

In 2023, "*Identification of data themes for the extensions of public sector High-Value Datasets*" [20] was published comprising of seven more categories to be potentially added to the existing HVD, namely: *(1) climate loss, (2) energy, (3) finance, (4) government and public administration, (5) health, (6) justice and legal affairs, (7) linguistic data*. In addition, some countries are considering or are already working on a determination of country-specific HVDs. The scientific literature supports these initiatives, and some of these studies are covered in the next section.

In general, the opening of HVD is an ongoing initiative in many countries within and outside the European Union, where initiatives implemented in EU Member States can be considered more harmonised and centralised, especially if they are launched as a response to the OD Directive call. However, the identification and opening of HVDs and the further assessment of their value can vary significantly, where approaches to identifying and assessing them are the most diverse. The next section examines some of these examples.

## 2.2 Research on High-Value Dataset opening

Although the topic of HVD is seen as having economic, technological and societal importance, it is relatively poorly represented in the literature (as also evidenced by a systematic literature review conducted by Nikiforova et al. [3]), where the few existing studies, can be classified into those (1) devoted to the identification of HVDs [5, 21], (2) assessing HVD compliance with technical requirements, including openness [4, 7], and the most represented category of studies that are (3) focused on assessing the value and impact of HVD opening [5, 6, 22, 23].

Studies devoted to the identification of HVDs are mostly oriented towards workshops or surveys/ questionnaires/ interviews [5, 21] found to be the most viable approach. For example, the study [5] aims to identify high-value datasets within a single case study **Latvia** by conducting a survey of individual users and SMEs as the OGD target audiences to assess their needs and satisfaction with the current data availability, identifying data sets that will provide greater value to users and businesses, and exploring how the availability of these data will impact their willingness to use OGD in the future. The survey findings indicate that these HVD would include data from *various registers* (statistics) and data on (1) *generally/globally topical subjects*, such as *COVID-19*, and (2) *local topics*, such as data related to *tourism in different country regions* (pointing out the interest in the local open data), *transport-related data with a focus on real-time data to track the transport location, accessibility of vehicles and places for people with disabilities, streets and traffic light data, radiation and noise levels*.

Most studies, however, focus on the analysis of the compliance of already published data to HVD (also ex-post impact assessment), or the general value of datasets to identify among them those of higher value. For example, Donker and Van Loenen [6] conducted a *Societal Cost-Benefit Analysis (SCBA)* of open HVD to examine the correlation between the associated with preparing data for publication as (linked) open data and the corresponding societal benefits by conducting a case-study of five HVDs in the **Netherlands**. Several scenarios were explored as part of the analysis, ranging from not publishing the dataset to publishing it as Linked Open Data (LOD). The findings revealed that, in general, the societal benefits of (linked) open data outweighs the costs, albeit this tends to vary depending on the datasets. In many cases, the costs of open data are an integral part of general data management costs, thus unlikely resulting in additional expenses. However, there are cases where the costs of preparing data for publication, including anonymization and aggregation, are high compared to the potential value of openly sharing the data. Nevertheless, the authors believe that, despite yielding a less favourable cost-benefits ratio for these datasets, the social benefits still is greater than without the open data version.

Stuermer and Dapp [22] employed a similar approach in their examination of the impact of available data by developing a framework based on the *SROI* concept *(Social Return On Investment)*. This framework was designed to assess the impact of open data, and it consists of four SROI values tailored to the open data context: (1) *input*, which encompasses resources such as money, personnel, infrastructure, equipment, and the additional resource of native data generated by the organisation (also "proprietary data"), (2) *output,* which refers to tangible deliverables, such as the establishment and operation of an open data portal, where data publishers share their data in appropriate technical formats, accompanied with metadata, ensuring their continuous maintenance and updates etc., (3) *outcome*, which encompasses all direct and indirect consequences of certain output actions of open data users, such as engineers, entrepreneurs, citizens, journalists, scientists, who utilise the available data in various forms, incl. hackathons, apps, new firms, data linking, research etc., and (4) *impact*, which represents the outcome adjusted for the effects that would have occurred in the absence of the intervention, and specifically addresses the value-creating consequences that are actually caused by the release of the data. These four values are then associated with the open datasets under consideration to develop a matrix of open data examples, activities, and impacts within each data category, which is expected to serve as an asset in evaluating the potential of dataset (HVD).

Alternatively, some studies assess HVD compliance with technical requirements, including openness. For example, Kević et al. [4] examines the openness and provision of geospatial data as data that can have a major impact on human activities, thereby considered HVD, in three South-Eastern European countries - **Croatia, Slovenia** and **Serbia** (with a particular focus on Croatia), using the revised Global Open Data Index (GODI) methodology they call *GODI Plus*. Focusing primarily on Croatia, they seek to understand *(1) how well do open data policy in Croatia support and promotes the provision of geospatial HVD, (2) whether and to what extent government agencies in Croatia are involved in the implementation of OD policies, (3) what are the best practices in Slovenia and Serbia that Croatia could learn from*. Their study found that Croatian open data policy is not sufficiently prescriptive, where there is a clear need to (1) encourage agencies to make their *data available* with a higher level of *openness*, (2) make *geospatial* HVDs more accessible for reuse, (3) improve *the usability and functionality of data portals* to enhance *public participation* and *collaboration*. Their results also indicate low data openness for Croatia and Serbia, with Slovenia considered an example in terms of published datasets and their compliance with a predefined list of geospatial HVDs, their *up to date'ness*, *metadata* and an *open license*, which, however, offers restricted access to data requiring prior registration. All three countries analysed demonstrated low levels of government engagement, which is a prerequisite for an effective and efficient OGD initiative.

Utamachant & Anutariya [7] evaluate and analyze HVD provided by **Thailand**'s OGD portal. Their analysis reveals issues in both - data provision and HVD compliance with technical requirements, including issues with *metadata* that significantly impact the *data quality* of HVD, which we found often underrepresented in the literature and practice. The study also highlights the importance of proper guidelines for data providers to follow in order to ensure the quality of the datasets published on the portal. For the latter, the study by Nikiforova et al. [3] can be mentioned, which, as a result of SLR on HVD conceptualised previous research in this area around four aspects (1) *data categories*, (2) *specific datasets*, (3) *data type*, (4) *data dynamism*. The authors first attempted to understand how HVD is defined and found that some studies define HVD according to the definition proposed in the Open Data Directive (EU) 2019/1024 [5], uses the "local" definition as is the case in Thailand relying on the HVD definition of the Electronic Government Agency of Thailand (EGA) combined with Government Open Data Index (GODI) and Open Data Barometer (ODB) [7], or use data categories defined in the G8 Open Data Charter [22]. Others use a high-level definition such as "*data that meet the actual needs*" [23], while some limit the definition to a specific area of interest such as business data [24] or disaster recovery planning [21], where the latter study defines the HVD as *critical datasets for their work, especially those that have a high granularity and spatial extent, with the potential for re-use*. In other words, there is no common and universal understanding of either definition or a list of HVDs, where, while a general belief in the importance of data in transforming it into value may be seen as a general component of the definition, various further characteristics - technical or legal - may be integral to it part. The latter - *data type* and *data dynamism*, in turn, cover the technical requirements that HVD are expected to meet in order to exploit its real value, where increased interoperability, including the provision of this data as *Linked Open Government Data (LOGD)* and the growing importance and value of *geospatial data, real-time data, sensor-generated data, citizen-generated data*, and data related to or sharing values with *Sustainable Development Goals* and the concept of *Smart Cities*, is recognized. Finally, the authors emphasise that the discovery and publication of HVD is a process that can be viewed as a life cycle and thus viewed as a holistic entity rather than disparate phases. This lifecycle has been compared to the *Plan-Do-Check-Act cycle (PDCA cycle)* or *Define-Measure-Analyse-Improve-Control phases of Lean Six Sigma*, which consists of at least: identifying potential HVDs, publishing the HVD, assessing the impact of the HVD and comparing it with the expected impact to decide whether adjustments are necessary.

In general, it can be said that most studies refer to *ex-post* analysis, i.e., analysis of HVD after the dataset has been published with limited research and, as a consequence, understanding of the entire process of HVD opening from their identification to their maintenance and impact measurement. Our study fills this research gap.

## 2.3 High-value datasets in Taiwan

The active promotion of OGD began in Taiwan in 2012 with the launch of the OGD portal, designed to integrate open data from various agencies and local governments. Although the portal has evolved over the years, including in terms of data availability with 47,000 datasets in 2020 [12] and more than 55 000 datasets in 2023, according to *Taiwan Open Government National Action Plan 2021-2024*, the private sector would prefer a more comprehensive system and mechanism to optimise the value of using government open data.

*Taiwan Open Government National Action Plan 2021-2024* [12], initiated by the National Development Council, encompasses five key commitments. The first and the core commitment - "*Complete Government Open Data and Data Sharing Mechanism*," focuses on the provision of high-value datasets. This initiative is strategically aligned with the goal of achieving the 16th Sustainable Development Goal (SDG) - "*Promote peaceful and inclusive societies for sustainable development, provide access to justice for all and build effective, accountable and inclusive institutions at all levels*", underscoring the plan's emphasis on sustainable development through open data and collaborative data sharing mechanisms. It consists of three commitments:

(1) to focus on **prioritizing high value data opening**, (2) to **strengthen data standards and format quality**, (3) to **establish processes for meeting public data needs**. These commitments are expected to (1) enhance governance transparency, (2) encourage value-added use of the civil sector, (3) implement public-private collaboration in the provision of innovative services. This is expected to, in turn, promote accountable government and good governance, open and transparent decision-making process to open data, working with the civil sector to create win-win scenarios. The Action plan identifies 3 milestones: (1) establish high-priority domain-specific open data areas, (2) advising agencies on the release of dynamic data in API format, being ongoing activities scheduled for the period between January 2023 and May 2024.

As part of the Action Plan, The Executive Yuan's Government Data Openness Advisory Group determined six (6) topics that represent the areas of HVD, namely: (1) **Climate and Environment**, (2) **Disaster Prevention and Relief**, (3) **Transportation and Transit**, (4) **Healthcare and Medical Services**, (5) **Energy Management**, and (6) **Social Assistance**. Within these categories, 20 sub-categories were identified, and another 408 specific datasets were identified as HVD and assigned to respective Taiwanese agencies to be opened in 2022. Currently, a total of five (5) ministries are expected to take respective actions of HVD opening - (1) Ministry of Environment, (2) Ministry of the Interior, (3) Ministry of Transportation and Communications, (4) Ministry of Health and Welfare, (5) Ministry of Economic Affairs, (6) Ministry of Health and Welfare. From the statistics collected by the respective agencies, it is known that only the Ministry of Environment is aware of applications developed using published HVDs - 35 - that fall under the topic of Climate and Environment. For other topics and datasets, respective data are not available. The reason for this is unknown, so we expect to gain relevant insight from the study on whether the subsequent use of opened HVD is monitored by responsible authorities, as well as how it is done.

## 3 Methodology

The semi-structured interview instrument was developed to collect empirical data to gain a deeper understanding of the current state of HVD in Taiwan by transforming the conceptual findings of our literature review into an interview protocol.

The interview contained HVD life cycle-related questions with which we delve deeply into (1) strategic, (2) technical, and (3) application aspects related to the topic in question. This includes examining selection criteria, revision cycles, incentives for promoting the propagation of high-value datasets, as well as the differences in requirements compared to general open data and evaluating the applications that emerge from these data sets. The interview questions are provided in Table 1. The choice of perspectives is justified by (1) the commitments of the *Taiwan Open Government National Action Plan 2021-2024* (see previous section) and (2) general phases/stages of the HVD lifecycle.

A purposive sampling approach was used to identify candidates for interview by approaching the agencies (listed by the Open Data Advisory Group of the Executive Yuan), which identifies HVDs and the government agencies responsible for opening them. Four (4) ministries - *(1) The Ministry of Digital Affairs (MoDA), (2) Ministry of the Interior (MoI), (3) Ministry of Transportation and Communications (MoTC), and (4) the Ministry of Environment, Department of Monitoring and Information (MoE),* comprise the sample for this study. Apart from the newly established *Ministry of Digital Affairs* (in 2022), other units have almost a decade of experience in promoting open data. The interviewees for this study include five (5) representatives from four (4) public agencies, who range from officer and section chief to director.

**Table 1.** Interview protocol

| Question |
|---|
| **1. Strategic Aspects** |
| *Q1a. How does your agency select topics, subcategories, and data sets for valuable data sets? What are the criteria and decisive factors for this, other than those defined by the Open Data Advisory Group of the Executive Yuan? Do you refer to any international practice or specific data openness goals?* |
| *Q1b. What is the established time cycle for implementing rolling changes as you refine topics, subcategories, and datasets for the HVD category?* |
| *Q1c. Are there any rewards or other incentives for agencies and peers involved in promoting valuable data sets?* |
| **2. Execution and Technical Aspects** |
| *Q2a. Besides public consultation, are there other methods or channels for needs assessment when selecting HVD?* |
| *Q2b. Do agencies have different requirements for valuable data sets compared to general open data? (for example, data quality, granularity, update frequency, integration methods, or data evaluation)* |
| *Q2c. Is there a plan to include non-personal data sets in valuable data sets?* |
| **3. Application Aspects** |
| *Q3a. Is there regular follow-up monitoring and analysis of applications and services developed from high-value datasets, e.g., in areas such as weather, traffic, net-zero initiatives?* |
| *Q3b. How do you measure the impact of an opened HVD (and whether this is done)? And how to evaluate this impact?* |

Two face-to-face interviews and two remote interviews were conducted (in line with the preferences of interviewees) in November 2023. Informed consent was obtained from the participants before commencing interviews. The duration of the interview ranged from 40 to 60 minutes. The interviews were documented through audio recordings and note-taking. Following the interviews, summaries were prepared by transcribing audio recordings, the accuracy of which was then confirmed by the interviewees. The list of interviewees and their schedule are presented in Table 2.

**Table 2.** Interviewee profiles and general interview data

| Service Unit | Participant and Interview Number | Interview Method |
|---|---|---|
| The Ministry of Digital Affairs (MoDA) | Section Chief (G1), Officer (G2) | In-Person |
| Ministry of the Interior, Department of Statistics (MoI) | Director (G3) | In-Person |
| Ministry of Transportation and Communications (MoTC) | Officer (G4) | Telephone |
| Ministry of Environment, Department of Monitoring and Information (MoE) | Section Chief (G5) | Online |

# 4 Results

In this section, we discuss the experiences of the ministries responsible for opening up HVDs in Taiwan, drawn from interviews, structured according to three perspectives - strategic, technical, and application, summarised in Table 3.

## 4.1 Strategic perspective of HVD opening

Taiwan's open data promotion strategy includes **a two-tier open data consultation system** and **a collaborative approach between the public and private sectors**. The first tier includes meetings of the Executive Yuan, and the second tier involves meetings of various ministries. At both levels, the inclusion of civil representatives is mandatory to ensure proper disclosure. Since the establishment of the *Ministry of Digital Affairs* in 2022, it has played a strategic role in promoting and advancing high-value datasets. Drawing inspiration from the G8 Open Data Charter, the Global Data Barometer, and the European Union's practices, the *Ministry of Digital Affairs* **(G1, G2)** proposed principles for listing High-Value Data Sets for other Ministries to follow. However, the Open Data Committee at the Executive Yuan level still needs to review these proposed datasets from various ministries. The interview results indicate that recent developments in data governance in the EU have influenced Taiwan's *Ministry of Transportation and Environment* **(G4)**, while the Ministry of the Interior **(G3)** primarily models its approach on Australia - *the First National Action Plan of the Australian Open Government Partnership 2016-2018* and their *Digital Marketplace* [25]. The *Ministry of Environment* **(G5)**, in addition to the EU model, draws inspiration from the US model.

Members of the two-tier Open Data Committee serve a two-year term, with the rolling review cycle for open data varying by agency, whereas the *Ministry of Digital Affairs* **(G1, G2)** revises annually, while the *Ministry of Environment* **(G5)** does it every six months. Other ministries do this irregularly.

The Ministry of Digital Affairs **(G1)** annually organizes the Open Data Quality Awards, which includes two categories: (1) data quality and (2) application cases. Currently, the awards for high-value datasets are integrated with the Open Data Quality Awards, without a separate mechanism for awarding individual units. Similar approach is followed by the *Ministry of Environment* **(G5)**. The *Ministry of the Interior* **(G3)** expressed **the need for financial support from the Executive Yuan for data collection, cleaning, and analysis** to facilitate the initiative across various ministries.

## 4.2 Execution and Technical Aspect of HVD

Ministries interviewed for the study indicated that they are collecting proposals and suggestions for open data and HVDs through **consultation meetings**. In addition, they incorporate **feedback from two online sources**: (1) **the open data platform** (https://data.gov.tw/) and (2) **the public policy participation platform** (https://join.gov.tw/). They also periodically collect data on the problems faced by their colleagues in their daily work.

All High-Value Data Sets on the OGD platform meet the **Gold Standard for data quality**, established by the *National Development Council*, according to which data must meet three standards: (1) **accessibility**, (2) **ease of processing**, and (2) **ease of understanding**. Accessibility refers to the validity of links to data resources and the ability for users to directly obtain data through these links without needing to log in or perform any additional actions. Ease of processing pertains to data being structured and available in an open format. Ease of understanding refers to the dataset metadata, including the 'encoding format' and 'main field descriptions', which should correspond to the fields represented in the data resources.

For HVDs compared to general open data, it is even more important to ensure that these data are provided in machine-readable form, complemented by access to them through API. Likewise, *dynamic data* and *geospatial data* are typically more commonly classified as HVD. However, in reality this is not always the case. HVDs in Taiwan include predominantly static data, such as domestic flight schedules, road safety training information provided by the *Ministry of Transportation*, living allowances for low and middle-income seniors provided by the *Ministry of the Interior*, national social welfare centre directories, and health promotion organisations for patients provided by the *Ministry of Health and Welfare*. Interviewed units noted that although these datasets are static, they have high download rates, suggesting potential for further development of services or applications. However, an interviewee from the *Ministry of the Interior* **(G3)** pointed out that HVD are called such not only because of the high download rates, but also because of the need to consider data trading issues, according to whom **"the real value of data is recognized when developers are willing to pay for it"**.

Finally, the issue of non-personal data was addressed. The open data platform **does not currently provide de-identified personal data**. The *Ministry of the Interior* **(G3)** on its official website offers only simulated data for approximately 2 million citizens. Access to de-identified personal data, such as health insurance and real estate transaction price registration, typically requires an application or payment. In response to the EU's advocacy of data altruism, the *Ministry of Digital Affairs* **(G1, G2)** is also working to promote guidelines to improve privacy and data for public good.

### 4.3 Application Aspects of HVD

In general, **tracking data reuse is a challenging task**. Since the principle of use of OGD by data users is free, with fees being the exception, ministries interviewed noted difficulties in tracking which private sector entities are using open data provided by government agencies.

**Table 3.** Interviewee summary

| Strategy/ministry | MoDA | MoI | MoTC | MoE |
|---|---|---|---|---|
| **Strategic perspective** | | | | |
| Q1a. dataset selection | 1) G8 Open Data Charter 2)Global Data Barometer 3) EU | Australia | 1) G8 Open Data Charter 2) Global Data Barometer 3) EU | 1) USA 2) EU |
| Q1b. time cycle for implementing changes | one year | irregularly | irregularly | six months |
| Q1c. rewards | ministry-level rewards, but not specifically tailored to HVD | ministry-level rewards, but not specifically tailored to HVD; suggests financial support from the Executive Yuan | ministry-level rewards, but not specifically tailored to HVD; has employee commendation and awards (within the ministry) | ministry-level rewards, but not specifically tailored to HVD |
| **Execution and Technical Perspective** | | | | |
| Q2a. needs assessment | two-tier Open Data Committee | two-tier Open Data Committee | two-tier Open Data Committee | two-tier Open Data Committee |
| Q2b. different requirements | Gold Standard and above | Gold Standard and above | Gold Standard and above | Gold Standard and above |
| Q2c. non-personal datasets | no | no, but have simulated data | no | no |
| **Application Aspect** | | | | |
| Q3a. regular follow-up monitoring and analysis | yes | yes | yes | yes |
| Q3b. impact evaluation | no | no | yes | part of it (AQI) |

The exception is the *Ministry of Transportation and Communications* (**G4**), which operates a separate open data platform called Transport Data eXchange (TDX). TDX uses a membership system, and while unregistered users can access and browse the system, registered users have more privileges, resulting in a significant number of users registering in it, which allows for tracking user access and further reuse of data. TDX deals with large volumes of dynamic, real-time data that are supported with API endpoints (connecting the data collection point to TDX and then providing access to the data from TDX to the service provider). In addition, due to the escalating load on servers, the *Ministry of Transportation and Communications* has implemented a membership system for distinguishing between users with high and low volumes of data use, which allows for more effective monitoring of data utilisation.

The *Ministry of Environment*, noting the diversity and amounts of environmental data, and the general interest in them, including their possible reuse, has placed particular emphasis on the application of air quality indexes (AQI) due to their relative ease of understanding and relevance to daily life. Interviewee **(G5)** noted that AQI data has the highest download rates among all environmental data.

Given the difficulties in tracking data reuse, ministries interviewed acknowledged that it is **difficult to accurately measure the impact of the data**. Although open data platforms annually announce benchmark reuse cases, most of them are the results of government agencies rather than private sector applications. Therefore, conducting standardised and systematic impact assessments is currently challenging.

# 5 Discussion

As data collection technologies and purposefulness evolve, the concept of High-Value Data Sets is undergoing significant expansion. This evolution ranges from definition that ranges from "*data that meet the actual needs*" [23] to specific categories that can be seen as advanced, such as *dynamic, sensor-generated data*, *citizen-generated data,* data related to or sharing values with *SDGs* and the concept of *Smart Cities* (EU). During interviews we identified that in Taiwan the value of data is determined through download rates or even trading potential, i.e., "*when developers are willing to pay for it*".

Although different ministries adhere to frameworks such as the G8 Open Data Charter, the Global Data Barometer, and EU policies, the wide range of HVDs inevitably leads to distinct interpretations, underscoring the rich diversity of HVDs and the variety of data purposes and collection methods across diverse themes. Reflecting this progress, the interviews revealed that Taiwan is preparing to augment its HVD portfolio in the second half of 2024 with three new categories: *(1) agricultural data*, *(2) geospatial data,* and *(3) government statistics*, marking a significant step in the data development and utilisation, as well as the entire open data ecosystem, taking a step closer to the EU HVD paradigm, where statistics and geospatial data are core HVD thematic categories, demonstrating the most progress (according to the ODMR2023). **Dynamic or real-time data**, which has proven to be of great importance at EU level, remains an untouched topic that requires further investigation, including incentivising discussions around it among ministries.

Addressing the diversity of data purposefulness, the government engaged public-private collaboration channels in the process of determining HVD themes. This approach involves private data developers and data managing authorities working together to propose data needs, aligning their respective objectives. This is expected to contribute positively to the identification and opening of the HVD. However, when HVDs are opened, the impact of their availability is expected to be measured as part of the HVD lifecycle. However, the **impact of open data is difficult to measure**, which is mainly due to the general paradigm underlying OGD, according to which the licence assigned to open data allows the data to be reused without having to declare it in the final product or service (open data are freely accessible and reusable), as also pointed out in [3]. This, along with the fact that most known open data reuses are the results of Taiwanese government agencies rather than private sector applications, stresses the need for establishing interaction and long-term cooperation with the open data ecosystem stakeholders with their prior identification, finding the way of their inclusion in this ecosystem, with appropriate preferably standardised and systematic mechanisms for user engagement and participation, which is found by the literature to be one of the key elements for a resilient and sustainable open data ecosystem (one of the recommendations by [26]). A feedback mechanism with the possibility of submitting re-uses on the OGD portal with their further promotion, as is done by the OGD portal of France, is one way to further explore this regard.

As in the EU context, **financial support** of data opening remains one of the critical barriers to the active, efficient, and effective high-value data management, as evidenced by the results of our interview, where the need for financial support from the Executive Yuan has been stressed. The tasks of collecting, cleaning, and analysing data mentioned by interviewees suggest further understanding of whether data literacy among government agencies responsible for data opening are sufficient, or rather they are expected to be supported, which is another critical barrier in the open data domain.

The **reward mechanism** that supports and incentivises data opening is also valid for HVD opening, where interviewees suggested a specific category associated with the HVD as an open data subdomain that should be added to existing reward categories.

# 6 Limitations

Although this study supplements the body of knowledge on non-EU countries being underrepresented in the given topic, there are some limitations that need to be addressed. First, we interviewed four of the five ministries responsible for opening HVD, which prevents us from fully generalising the results to the Taiwanese context, suggesting that interviewing the last ministry would reveal more shortcomings allowing us to come with a final and more comprehensive list and respective recommendations.

Second, in this study, we only considered the perspective of government officials, not including scholars, experts, or data developers who could provide a more diverse and complete understanding of the current state of affairs of the HVD in Taiwan. However, the purpose of this study was to understand the HVD opening lifecycle, where government officials are responsible for this and constitute therefore the audience of interest to establish this understanding. In the future, expanding the scope of the study, it would be relevant and useful to consider other stakeholders of the open data ecosystem, considering the quadruple helix, i.e., including both policy, academia/science, industry/business, and society/ citizens. Industry and society helixes with data developers may be of greater interest as main data reusers, however, their involvement may pose a challenge, which is due to the limited understanding of data developers with HVD experience and the lack of representative successful case studies of HVD reuse in Taiwan. This confirms the need to create a mechanism to identify and promote HVD reuses.

# 7 Conclusions

This study offers a unique perspective on High-Value Dataset research from Taiwan, a non-EU country renowned as a global leader in ICT production. This approach enriches the national dimension of HVD research. Through secondary data analysis and interviews, the study delves into the strategies and initiatives of four key ministries in promoting HVD - *(1) The Ministry of Digital Affairs, (2) Ministry of the Interior, (3) Ministry of Transportation and Communications, and (4) the Ministry of Environment, Department of Monitoring and Information*. By conducting semi-structured interviews with these entities, we delve into (1) *strategic* aspects related to the HVD determination process, (2) *execution and technical* aspects, and (3) *application* aspects, identifying some good practices and weaknesses to be further examined and fixed.

From a macro policy planning perspective, we find that there is a need for more national-level perspectives to enrich research on the HVD ecosystem, for instance, in Australian settings, which serves as a model for other countries due to its good start in HVD, with significantly fewer details about the further progress. Secondly, from a cross-organizational communication perspective, increasing data interoperability by enabling data connectivity, such as linked open data, is critical to maximise their value. Unfortunately, our interviews did not reveal examples of such initiatives.

Future studies could delve deeper into the dynamics of cross-ministerial collaboration and the best practicality of HVD application. Finally, from a micro policy implementation perspective, this study observed that the Ministry of Digital Affairs staff responsible for open data management are not the same people who previously performed these tasks. The concern is whether this change impacts the promotion of high-value data leading to discontinuities in policy, or indirectly impacts policy due to differences in staff digital literacy.

As the next step, considering the popularity of HVD topic and practical advances in this area within the EU and its Member States, our future research will take a holistic comparative exploratory case study approach, where Taiwanese approach will be compared and learnt lessons will be exchanged with the EU Member State - Estonia that according to the recent Open Data maturity Report [19], is significantly ahead of other EU Member States in implementing Commission Implementing Regulation (EU) 2023/138. With this, we expect to gain a more understanding and contribute to a more holistic HVD life cycle, understanding best practices from both Taiwan - recognized for its achievements in ICT and open data development, and Estonia - a pioneer in digital development among EU countries and especially digital government, where in the upcoming years, the development of the OGD initiative (including, HVDs and OGD portal contexts) is planned. This is expected to contribute positively to a general open data ecosystem within the EU and outside it, contributing to more interoperable and sustainable open data ecosystem development and maintenance.